\documentclass[fleqn,usenatbib]{mnras}

\usepackage{newtxtext,newtxmath}

\usepackage[T1]{fontenc}
\usepackage{ae,aecompl}

\usepackage{graphicx}	
\usepackage{amsmath}	
\usepackage{amssymb}	

\title[Multiwavelength timing analysis of GX 339-4 ]
{Characterization of the Infrared/X-ray sub-second variability for the black-hole transient GX 339-4}
 
\author[F. M. Vincentelli et al.]{F. M. Vincentelli,$^{1,2,3}$\thanks{E-mail: federicomaria.vincentelli@oa-roma.inaf.it} P. Casella,$^3$ T. J. Maccarone$^4$, P. Uttley$^5$, P. Gandhi$^6$,\newauthor T. Belloni$^2$, B. De Marco$^7$, D. M. Russell$^8$, L. Stella$^3$, K. O'Brien$^9$
\\
$^{1}$DiSAT, Universit\'{a} degli Studi dell'Insubria, Via Valleggio 11, I-22100 Como, Italy\\
$^{2}$INAF, Osservatorio Astronomico di Brera Merate, via E. Bianchi 46, I-23807 Merate, Italy\\
$^{3}$INAF, Osservatorio Astronomico di Roma, Via Frascati 33, I-00078 Monteporzio Catone, Italy\\
$^{4}$Department of Physics \& Astronomy, Box 41051, Science Building, Texas Tech University, Lubbock TX, 79409-1051, USA\\
$^5$Anton Pannekoek Institute for Astronomy, University of Amsterdam, Science Park 904, 1098XH Amsterdam, Netherlands\\
$^6$Department of Physics and Astronomy, University of Southampton, Southampton SO17 1BJ\\
$^7$Nicolaus Copernicus Astronomical Center, Polish Academy of Sciences, Bartycka 18, PL-00-716 Warsaw, Poland\\
$^8$New York University Abu Dhabi, PO Box 129188, Abu Dhabi, UAE\\
$^9$Department of Physics, Durham University, South Road, Durham, DH1 3LE, UK
}

\date{Accepted XXX. Received YYY; in original form ZZZ}

\pubyear{2017}

\begin{document}

\label{firstpage}
\pagerange{\pageref{firstpacmge}--\pageref{lastpage}}
\maketitle

\begin{abstract}
We present a detailed analysis of the X-ray/IR fast variability of the Black-Hole Transient GX 339-4 during its low/hard state in August 2008. Thanks to simultaneous high time-resolution observations made with the VLT and RXTE, we performed the first characterisation of the sub-second variability in the near-infrared band -- and of its correlation with the X-rays -- for a low-mass X-ray binary, using both time- and frequency-domain techniques. We found a power-law correlation between the X-ray and infrared fluxes when measured on timescales of 16 seconds, with a marginally variable slope, steeper than the one found on timescales of days at similar flux levels. We suggest the variable slope -- if confirmed -- could be due to the infrared flux being a non-constant combination of both optically thin and optically thick synchrotron emission from the jet, as a result of a variable self-absorption break. From cross spectral analysis we found an approximately constant infrared time lag of $\approx0.1$s, and a very high coherence of $\sim 90$ per cent  on timescales of tens of seconds, slowly decreasing toward higher frequencies. Finally, we report on the first detection of a linear rms-flux relation in the emission from a low-mass X-ray binary jet, on timescales where little correlation is found between the X-rays and the jet emission itself. This suggests that either the inflow variations and jet IR emission are coupled by a non-linear or time-variable transform, or that the IR rms-flux relation is not transferred from the inflow to the jet, but is an intrinsic property of emission processes in the jet.
\end{abstract}

\begin{keywords}
black hole physics-- X-rays: binaries.-- relativistic jets -- stars: individual: GX 339--4 
\end{keywords}



\section{Introduction}
Black hole X-ray transients (BHT) are a class of low-mass X-ray binaries (LMXRBs) in which long periods of quiescence are interrupted by dramatic X-ray outbursts; strong activity is also present at longer wavelengths, from ultra-violet (UV) down to radio frequencies. The overall emission from these objects is interpreted as the result of three main emitting components, whose properties and relative contribution to the broad-band emission vary along the outburst: a thermally emitting, optically thick, geometrically thin accretion disc, a Comptonizing hot plasma, and a collimated jet. {Emission from the donor star instead is usually found to be faint and negligible during the outburst.}

During the hard state (i.e. when the Comptonizing medium dominates the X-ray emission) a strong radio flux with a usually flat or slightly inverted spectrum, which can extend up to the infrared (IR) band, is also detected \citep{tannabaum1972,fender2001}. According to the \citet{blandford1979} model, the superposition of self-absorbed synchrotron emission profiles coming from different regions of a relativistic jet can reproduce the observed spectrum at longer wavelengths. This model is based on the fundamental assumption that the energy of the electrons is continuously replenished, in order to balance the radiative and adiabatic energy losses. The mechanism providing such energy remains unclear, although a few ideas have been proposed. A possible solution could come from the conversion of Poynting flux into internal energy through magnetic reconnection \citep{lyubarsky2010}. An alternative solution involves internal-shocks, by analogy with the model originally proposed to explain the variability observed in Gamma-Ray Bursts \citep{kobayashi1997} and in blazars \citep{ciao}. This idea has been applied recently to the less-energetic case of jets in BHTs \citep{jamil2010,malzac2013}, with encouraging results from both the spectral and the timing point of view. According to these models, shells of plasma are continuously ejected into the jet with variable velocities. Due to the difference of velocity, the shells eventually collide and merge: the resulting shocks can convert part of the differential kinetic energy of the shells into internal energy, re-accelerating the electrons so as to -- at least partially -- balance the energy losses. A key prediction is the presence of strong variability in the optical-infrared (OIR) emission from the jet, also on sub-second timescales, plausibly correlated with the variability observed in the X-rays from the inflow  \citep{malzac2014,drappeau2015}.

The connection between the inflow and the jet in BHTs has been investigated by different authors, with a growing number of studies focussing on the properties of the OIR variability of XRBs and its correlation with the X-rays \citep{mirabeletal1998,eikenberry1998,kanbach2001,spruit2002}. A turning point in this respect was the discovery, in the early 2000s, of the existence of correlations -- on timescales of $\approx$ hours -- between the X-ray and the radio and/or IR luminosity during the hard state \citep{hannikainen1998,corbel2003,gallo2003}. In the same years, the first detailed characterisation of the fast X-rays/optical-UV variability was obtained for the BHT XTE J1118+480 (\cite{kanbach2001} and \cite{malzac2003} for the optical, and \cite{hynes2003} for the UV). All the measured cross-correlation functions (CCFs) indicated that the X-ray radiation led the O-UV variability, showing, however, also complex trends. An interesting feature was measured in the optical/X-ray CCF, where a small anti-correlation at negative lags was found: this was interpreted in terms of the jet (emitting in the optical) and the accretion flow being powered by the same magnetic energy reservoir \citep{malzac2004}. Such a model also reproduces the observations made by \cite{hynes2003}, if a major contribution from reprocessing is also assumed to be present in the UV. 


After the identification of an optical/X-ray subsecond lag in GX 339-4 during the 2007 outburst \citep{gandhi2008}, a further, complete characterisation of the optical/X-ray variability was then made by \cite{gandhi2010}. The optical power spectrum showed a complex broad-band noise structure, with a Quasi Periodic Oscillation (QPO) at $0.05$ Hz (while, there was no clear evidence of QPO in the X-rays). Such complexity was reflected also in the correlation between the two bands. The CCF showed a sharp asymmetric peak at small lags, and a broader structure for lags > 10 s. In the frequency domain, the time lags showed that the optical emission lagged the X-rays by $\approx 10$s on long timescales; a constant optical lag of $\approx 0.1$ s was instead present on shorter timescales ($\nu>0.1$ Hz). The authors interpreted the optical emission on long timescales in terms of reprocessing from the outer disc, and claimed the presence of an optically emitting relativistic jet for the fast variability.


Such a complex observational picture makes it very difficult to have a clear idea of the origin of the emission in the Optical/UV. This puzzling phenomenology probably is due to the fact that there are several spectral components that can emit in those bands. This complication is less important at longer wavelengths, where the jet is expected to dominate the emission. It was indeed thanks to fast-photometry observations performed in the IR band, that the first unambiguous presence of a rapidly varying jet emission was found in GX 339-4 \citep{Casella2010}.
During the low-hard state of the source these authors found intrinsic variability up to 8 Hz, strong enough to be inconsistent -- for brightness-temperature reasons -- with any kind of thermal emission. Moreover through simultaneous observations made with the Rossi X-ray Timing Explorer (RXTE), a strong correlation between the IR and the X-ray light curve was measured with an IR lag of $\approx 100$ ms. This was interpreted within a scenario in which the variability propagates from an X-ray emitting region (either the hot inflow or the base of the jet) along the jet and then re-emitted as synchrotron radiation (it is unclear whether this is optically thin or thick emission). The recent discovery of a $\approx$0.1 s optical/X-ray lag in the BHT V404 Cygni \citep{gandhi2017} is a clear confirmation of this scenario, which also allows to put constrains on the physical scale over which plasma is accelerated in the inner jet.

In this paper we present a more detailed characterisation of the X-ray/IR variability of the same data set used by \citet{Casella2010}, in order to better understand the way in which the variability is transferred along the jet, and investigate what is the jet emission mechanism. The paper is organised as follows: after a description of the data set, in Section \ref{flux} we present X-ray/IR flux-flux diagram computed at high time resolution (16~s); in Section \ref{fourier} we show the results of Fourier analysis (including cross-spectral techniques), while in Section \ref{rms} we report the first rms-flux relation computed in the IR band. Possible physical implications are discussed in Section \ref{discussion}.

\section{Observations}

\begin{figure}
  	\includegraphics[width=\columnwidth]{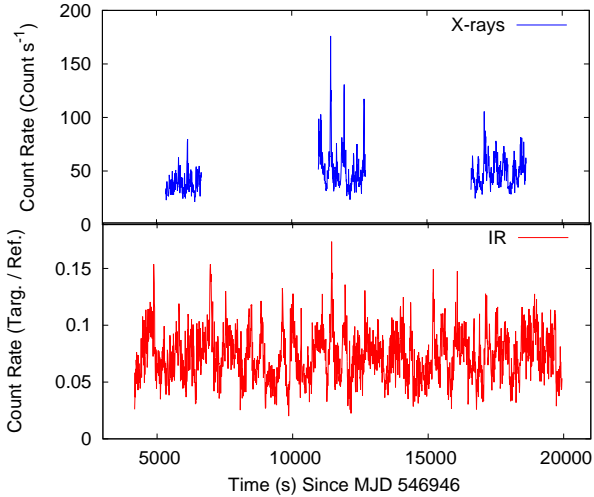}
  	\caption{
 	{ {\it Upper panel}: X-ray light curve of GX 339-4, as observed with RXTE. The two long gaps in the data are due to Earth occultation of the source. \it Lower panel}: Simultaneous Infrared light curve, as observed with ISAAC@VLT. Both light curves have been rebinned as to obtain a time resolution of 8 seconds. 
 	}
  	\label{fig:ltrcv}
  \end{figure}

The analysis carried out for this work is based on X-ray and IR simultaneous observations of the BHT GX 339-4 carried out on August 18 2008 (MJD 54696), during the beginning of the decay of a short-lived hard-state-only outburst \citep{russell2008,kong2008}. The Rossi X-ray Timing Explorer (RXTE) satellite performed simultaneous observations in the X-rays for 3 consecutive satellite orbits, for a total of $\approx$ 4.6 ks. Three intervals (corresponding to the periods in which RXTE could observe the source) of simultaneous X-ray and IR data were obtained (Fig. \ref{fig:ltrcv}).

\subsection{Infrared data}
The IR data were collected with a $\approx$ 16 ks long fast photometry observation, made with the Infrared Spectrometer And Array Camera (ISAAC), mounted on the 8.2-mUT-1/Antu telescope at ESO's Paranal Observatory \citep{Moorwood 1998}. The data were collected using the \emph{FastPhot} mode in $K_S$ band (central wavelength: 2.16 $\mu$m; width: 0.27 $\mu$m). The integration time (DIT) of the observation was 62.5 ms. Each frame was stored and stacked in a \lq\lq data-cube" of 2500 slices (corresponding to segment of IR data of 156.25 seconds). The length of the gaps between each cube is 3 second long.

The field of view (FoV) consisted of a 23'' $\times$ 23'' square, approximately centred on the target GX 339-4, and included a bright reference star ($K_S$ = 9.5), and a fainter comparison star ($K_S$ = 12.8), located respectively 13.6 arcsec South and 8.9 arcsec North-East of the target.

The light curve extraction was done using the ULTRACAM data reduction software tools \citep{dhillon2007}. Standard aperture photometry was performed, using the parameters derived from the bright reference star position and profile.
To account for seeing effects, the ratio between the source and the bright reference star count rate was used. The mean count
rate from the reference star light curve was $F_{ref}$ = $376672 \pm 54$ counts s$^{-1}$ , while the mean ratio of target to reference star flux was $\sim 0.07$. Times were then put in Barycentric Dynamical Time system of reference.

\begin{figure}
 	\includegraphics[width=\columnwidth]{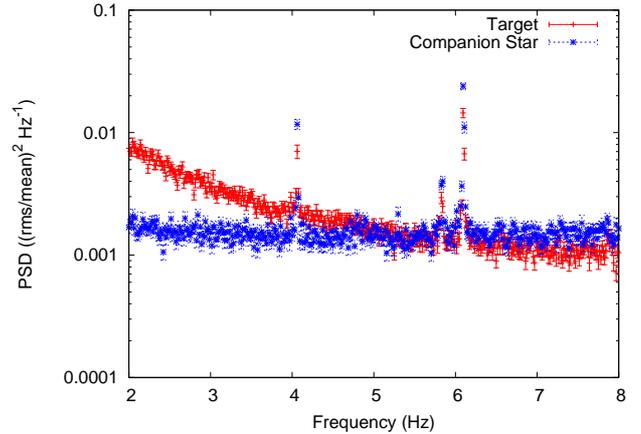}
 	\caption{
	 Section of the Fourier power spectra of the Infrared light curve of the target and the comparison star, showing the three instrumental peaks in both signals. The constant noise level was not subtracted.}
 	\label{fig:psd_zoom}
 \end{figure}

  \cite{Casella2010} noticed that the IR power spectrum presented some instrumental features at $\nu\approx$ 6 Hz, which can affect the high frequency measurements. To confirm this, we computed the target's and the comparison star's power spectrum at high frequencies ($\nu$>2 Hz).  Fig \ref{fig:psd_zoom} shows three strong peaks respectively at $\approx$ 4.1, 5.9 and 6.1 Hz present in both sources, proving that the origin of the peaks is instrumental. Nonetheless their effect on the correlation with the X-rays is negligible as such signals are present only in the IR band.

\subsection{X-ray data}
We used data from the Proportional Counter Array (PCA) on board of RXTE, analysing the observations corresponding to the three orbits. We used PCA \lq\lq Binned mode" observations with 7.8125 ms time resolution. Only 2 proportional counter units (PCU) were active during the whole observation. Standard HEADAS 6.17 tools were used for data reduction. The X-ray light curve was extracted in the 2-15 keV (channels 0-35) energy range. A background light curve with 16s of time resolution was estimated using the FTOOL \emph{pcabackest} (the background for RXTE PCA observations is not directly measured, but it is estimated by a model, with a minumum time integration of 16 s). The background curve was resampled at 7.8125 ms and subtracted from the extracted lightcurve.  As our observations had a mean count rate of 23.5 counts $s^{-1}$ $PCU^{-1}$, we used the RXTE PCA faint background model (for net count rates $\leq$ 40 counts s$^{-1}$ PCU$^{-1}$): the average background value was 19$\pm$ 2 count s$^{-1}$. The barycentre correction for Earth and satellite motion was applied using the HEASARC FTOOL \emph{barycorr} which converts times into the Barycentric Dynamical Time system of reference. The light curves were then rebinned at 0.0625s so as to have the same time steps as the IR curve. 
\section{Data Analysis}

\subsection{Fast flux-flux correlation}
\label{flux}
The discovery of a strong power-law correlation between the radio and X-ray bands during the low-hard state \citep{corbel2003,gallo2003}, as well as between the OIR and X-ray bands \citep{homan2005,russell2006,coriat2011} opened a broad discussion on the physical implications of the behaviour shown by different sources on the radio/X-ray and OIR/X-ray planes. {However, even though many studies focused on the strict simultaneous X-ray/radio properties (e.g. \cite{mirabeletal1998,klein-wolt2002,wilms2007}), most of the past studies on the flux-flux relations were made by using quasi-simultaneous observations; i.e. by averaging non-simultaneous measurements made (generally) within one day. }Here we present for the first time the IR/X-ray correlation diagram studied using high time resolution, strictly simultaneous observations.

The correlation diagram with full time resolution turned out to be too noisy to display any trend; therefore we re-binned both light curves with a 16-second time resolution. The errors were estimated with the standard propagation formula. 
 
\begin{figure}
  	\includegraphics[width=\columnwidth]{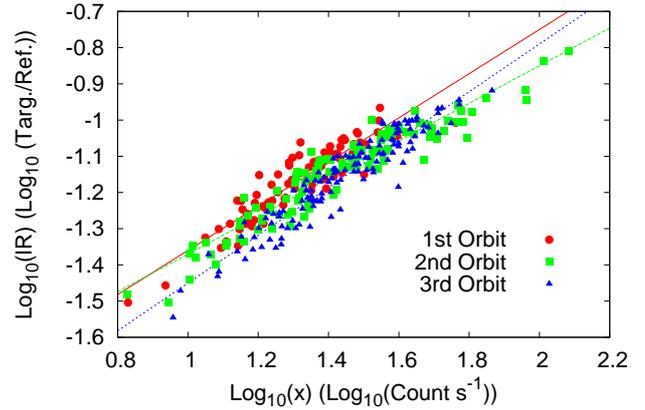}
  	\caption{Flux-flux diagram with 16 s of time resolution for the three orbits. The Targ./Ref IR light curve is plotted as a function of the X-ray count rate. Red circles, green squares and blue triangles correspond respectively to the first, second and third orbit. Continuous, dashed and fined dashed lines correspond to the best fit for first, second and third orbit respectively.}
  	\label{fig:flux-flux}
  \end{figure}

\begin{figure}
  	\includegraphics[width=\columnwidth]{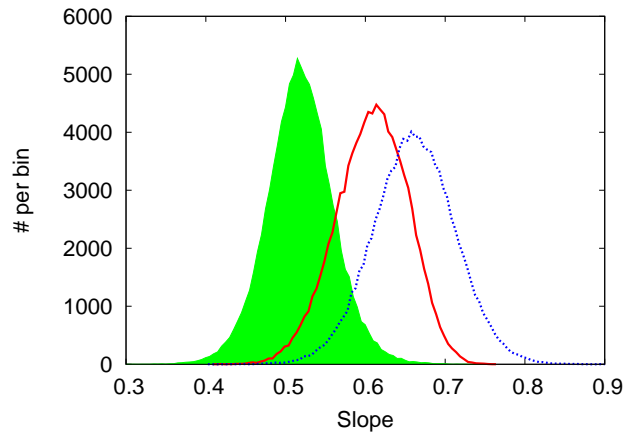}
  	\caption{Distribution of the simulated slopes after N=$10^5$ different sub-sampling.  Red, green and blue histograms are respectively the slope distribution for the first, second and third orbit.}
  	\label{fig:hist-slope}
  \end{figure}

The resulting correlation diagram for the three separate orbits is shown in Fig. \ref{fig:flux-flux}: a power law-like trend is clearly visible. This is the first evidence for an X-ray-IR power law dependence on such short time scales. To quantify the correlation, we fitted the data -- using the Ordinary Least Squares method -- assuming that $F_{IR}=a\cdot F_{x}^b$ and therefore $log_{10}(F_{IR})=b\cdot log_{10}(F_{x})+log_{10}(a)$, obtaining an average slope of $\sim 0.55$. As the statistical uncertainties are much smaller than the observed scatter, $\chi^2$ test cannot be applied to properly quantify the observed trend. We therefore obtained a more accurate estimate of the slope and its uncertainty applying a bootstrap method to each orbit separately. Operatively, we extracted N times a sub-sample of M different couples of points from the X-ray/IR simultaneous 16-second light curve of each orbit, in order to have N slopes. From the distributions obtained assuming M=80 and N=$10^5$, we then extracted the average values for both slopes and intercepts, their variance providing an estimate of their uncertainties. The obtained values are reported in Table \ref{tab:fit}.  As we can see also from Fig. \ref{fig:hist-slope}, there is marginal evidence for a different slope between the 2nd and the 3rd orbit (2.8 $\sigma$) and between the 1st and the 2nd (2.1 $\sigma$).  
 
 \begin{table}
 \caption{Parameters of the linear fit ($log_{10}(IR)$=b*$log_{10}$(X)+c) estimated for the flux-flux diagram for each of the orbits separately.}
 \label{tab:fit}
 \begin{tabular}{ccc}
  \hline
  Orbit &  b & c\\
  \hline
  1st & 0.61 $\pm$ 0.03 & -1.97 $\pm$ 0.05\\[2pt] 
  2nd & 0.52 $\pm$ 0.03 & -1.89 $\pm$ 0.05\\[2pt]
  3rd & 0.66 $\pm$ 0.04& -2.11 $\pm$ 0.04\\[2pt]
  
  \hline
 \end{tabular}
 \end{table}

\subsection{Fourier analysis}
\label{fourier}
 Fourier domain analysis techniques are very useful for characterising the X-ray variability properties of LMXRBs. In particular a very powerful frequency-domain tool is cross-spectral analysis, which enables the lag behaviour on distinct time-scales of variability to be much more cleanly separated than with time-domain methods such as the CCF \citep{nowak1999,uttley2014}. The application of these techniques on fast multi-wavelength observations can help to probe the disc/jet connection down to short timescales. {While all past studies were done using optical data \citep{malzac2004,gandhi2010,veledina2015}, we present the first full cross-spectral characterisation using X-ray and IR high time resolution observations of an XRB.}

The maximum frequency for the Fourier analysis ($\nu_{max}$=8 Hz) is set by the IR time resolution (0.0625 s). To improve the statistics, the analysis was carried out by dividing the time series into segments of 1024 time bins (i.e. 64 s), averaging them to obtain the final power(cross)-spectrum\footnote{The CCF measured by \citet{Casella2010} showed a width of $\approx$ 20 s, therefore, a 64 s long segment is sufficient to measure the cross spectrum without truncating the CCF and therefore allows lags to be estimated safely, avoiding bias effects \citep{jenkins,alston2013}.} \citep{vanderklis1989}. The choice of the segment length is dictated by the presence of gaps in the IR light curve. A further geometric re-binning in frequency was applied to increase the statistics at high-frequencies. The power spectral density (PSD) for the X-ray and the IR light curve was calculated in units of squared fractional rms per Hz \citep{belloni1990,vaughan2003}.  It is possible to show that if the effective sampling time $T_{\rm samp}$ is less than the nominal time bin $T_{\rm bin}$, the Poissonian noise level changes by a factor $T_{\rm samp}/T_{\rm bin}$ (\cite{vaughan2003,gandhi2010}). The ISAAC detector, in the configuration used during these observations, had a read out time of 10 ms \citep{Casella2010}, which means that $T_{\rm samp}/T_{\rm bin}=0.84$. This corrective factor was applied to the Poisson noise evaluation.
 
\begin{figure}
 	\includegraphics[width=\columnwidth]{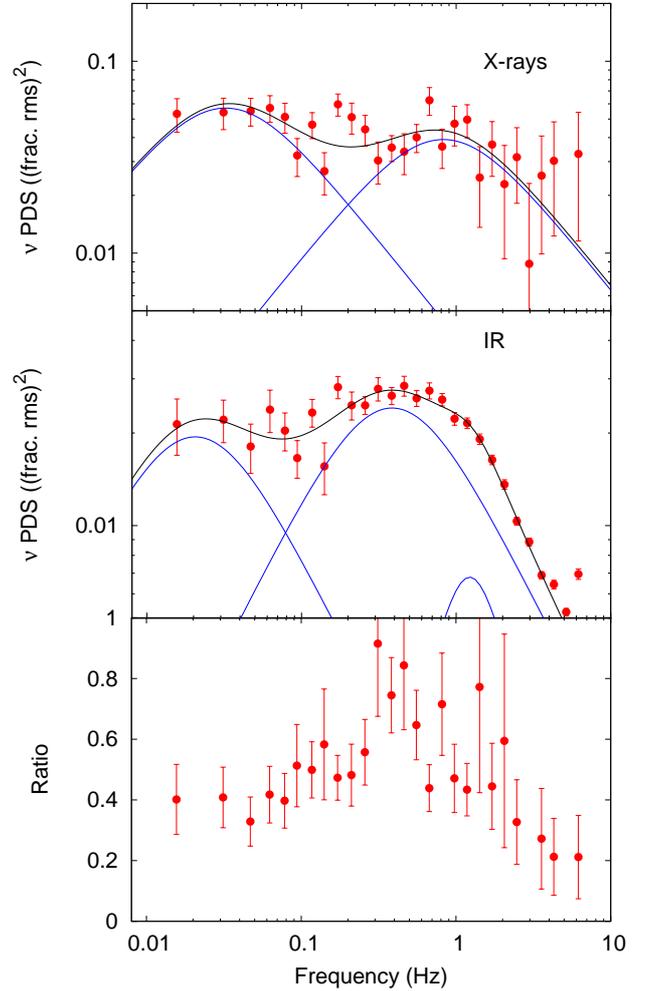}
 	\caption{
	{\it Upper panel:} X-ray power spectral density  data and best fit  using 1024 bin/segment and a geometric (i.e. multiplicative) rebinning factor of 1.2. The blue continuous lines represent the Lorentzian components obtained by the fit, while the black line is the full model.
    	{\it Central panel:} IR power spectral density data and best fit using 1024 bin/segment and a geometric rebinning factor of 1.2. 
	{\it Lower panel:} Ratio between IR and X-ray power spectrum. The plot shoes clearly that the IR PSD has relative increase followed by a break at $\approx$ 1 Hz.}
 	\label{fig:psd}
 \end{figure}

The noise-subtracted PSDs are shown in Fig. \ref{fig:psd}, no evidence of Quasi Periodic Oscillations is found. The highest frequency bin of the IR power spectrum has a significant deviation from the general trend due to the instrumental peaks found at $\approx$ 6 Hz.  The X-ray power spectrum can be reasonably well described with a power law with a slope of $1.1\pm0.1$ (giving a reduced $\chi^2$ of 1.5 with 26 degrees of freedom), while a broken power law does not improve the fit significantly. The IR power spectrum requires a more complex model, a broken power law ($\beta_{\rm low}\sim 1$, $\nu_B\sim 1$ Hz, and $\beta_{ \rm high}\sim2$) giving a reduced $\chi^2$ of 2.4 with 21 degrees of freedom.

To characterize better the two PSDs, we fitted them using Lorentzian profiles \citep{belloni2002}.  IR-band, data above 4 Hz were excluded due to the instrumental features (see Fig. \ref{fig:psd_zoom}). Two components were sufficient to describe the X-ray PSD, while a third Lorentzian component was necessary for the IR band (see Tab. \ref{fit_powe}). We note that, while the lowest-frequency component appears to be the same in both PSD (i.e. with the same centroid and width), none of the other two components in the IR PSD seems to match the high-frequency component in the X-ray PSD. We explored this issue further, trying to fit the IR PSD by fixing the parameters of the two X-ray components (but allowing their normalisation to vary), and adding a third Lorentzian component. However, such a model did not manage to reproduce the data, returning a reduced $\chi^2 > 2$.
 
\begin{figure}
	\includegraphics[width=\columnwidth]{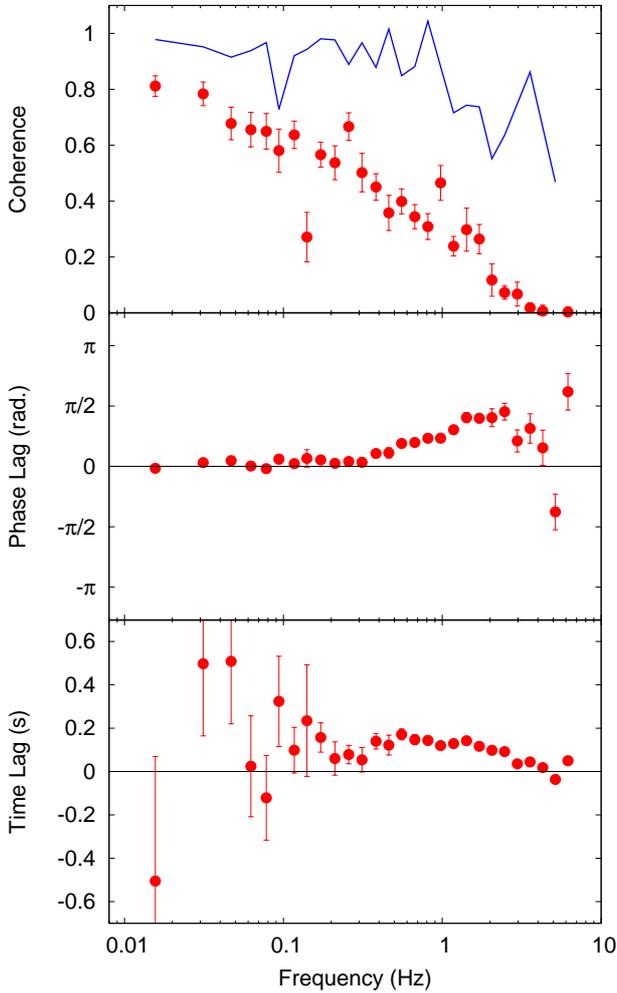}
	\caption{\emph{Upper Panel}: intrinsic coherence between the X-ray and the infrared light curves. Blue line represents the expected coherence for two perfectly coherent signals with same PSD shape and statistics as in our dataset. \emph{Middle Panel}:  Phase lags between the X-ray and the infrared light curve. Positive lags mean the infrared variability lags the X-ray variability. \emph{Lower Panel}: time lags between the X-ray and the infrared light curve. All are computed from 1024 bin per segment with a logarithmic binning factor of 1.2}
	\label{fig:lags}
\end{figure}

%
%
%
%
%
%
%
%
%
%
%
%
%

 \begin{table}
 \label{fit_powe}
 \centering
 \caption{Parameters from the fit with multiple Lorenztian component. The fitted function is the following: $PSD(\nu)=\frac{r^2 \cdot \Delta}{\pi} / [(\nu-\nu_0)^2+\Delta^2]$. When the error is not present, the parameter has been fixed. }
 \label{tab:fit_lorenzt}

\begin{tabular}{ccccccc}
  \hline
\# Comp. & $\nu_0$ & $\Delta$ (Hz) & r  \\  
  \hline
\\
X-rays & & \multicolumn{2}{c}{($\chi^2/dof$ = 30/24)}\\
  \hline
1 & 0 &   $0.03\pm0.01$& 0.29$\pm$0.02\\[2pt]
\\
2 & 0 &  $0.83^{+0.01}_{-0.02}$ & $0.24\pm0.01$ \\[2pt] 
\\
\\
IR & & \multicolumn{2}{c}{($\chi^2/dof$ = 22.6/18)}\\
  \hline
1 & 0 & 0.02$\pm$0.01 &0.17$\pm$0.01 \\[2pt]
\\
2 & 0    & 0.39$^{+0.09}_{-0.05}$ & 0.19$\pm$0.01  \\[2pt]
\\
3 & $0.98^{+0.33}_{-0.37}$   & 0.72$^{+0.08}_{-0.22}$ & 0.12$\pm$0.03 \\[2pt]
  
  \hline
 \end{tabular}
 \end{table}



From the cross spectrum  two quantities are derived that give information on the causal relation between the two signals: the coherence and the lags. The former gives a measure of the degree of linear correlation of two signals \citep{vaughan1997,nowak1999}, while the latter measures for each frequency component their relative shift in phase (or in time). Coherence and lags were computed following the recipe in \cite{uttley2014}. 
 
The intrinsic coherence for the X-ray and IR data is shown in Fig. \ref{fig:lags}. A high level of coherence is present on long timescales. From a value of almost 0.9 at $\approx$ 0.01 Hz, the degree of coherence declines to $\sim$0.2 at $\approx$1-2 Hz, while at the highest frequencies no significant degree of correlation is detectable. We verified whether the drop in the coherence could be due to bias effects. We therefore simulated two perfectly coherent signals with the same PSD shape and statistics as in our dataset and then evaluated the coherence. The resulting coherence (Blue line, Fig. \ref{fig:lags}: Top panel),is significantly higher than the measured one, showing that bias effects cannot be responsible for the observed decreasing trend.

Phase and time lags are also shown in Fig. \ref{fig:lags} (middle and lower panel, respectively). The phase lags show an increasing trend (considering the absolute value) between 0.1 and 2 Hz, corresponding to a roughly constant time lag of 100 ms, consistent with the lag found with the CCF by \cite{Casella2010}.

\subsection{rms vs. flux relation}
\label{rms}

A linear relation between the absolute rms and the flux has been reported in all classes of accreting sources: XRBs, Active Galactic Nuclei \citep{uttley2001} , Ultra-luminous X-ray Sources \citep{heil2012}, White Dwarfs \citep{scaringi2012} and more recently also in Young Stellar Objects \citep{scaringi2015}. Given also the presence of a lognormal distribution of the fluxes, it is usually assumed that this phenomenon arises from the multiplicative coupling of the fluctuations which propagate inwards in the accretion inflow \citep{uttley2005}. Recent studies have also found this relation in the variable emission at wavelengths where the accretion inflow is not thought to be the dominant emitting component, raising questions as to the nature and origin of this property \citep{gandhi2009,edelson2013}. In order to fully characterise the IR variability in our dataset, we also checked for the presence of a correlation between rms and flux in the IR light-curve.

The computation required several steps. First, the light curve was divided in K segments in order to have K power spectra (each with a fractional squared rms normalisation). Then, the segments were re-ordered in count rate.
Then the power spectra and the flux levels of each group of N adjacent segments were averaged together, resulting in K/N points in the rms-flux plane. 
For consistency with previous studies, the rms was normalised in absolute units, and calculated over relatively narrow frequency ranges, each with a lower boundary corresponding to the length of the segments. Thus, 1-second long segments were used to compute the rms-flux relation at frequencies above 1 Hz, 4-second long segments were used to compute the rms-flux relation at frequencies above 0.25 Hz, and so on.

\begin{figure}
 	\includegraphics[width=\columnwidth]{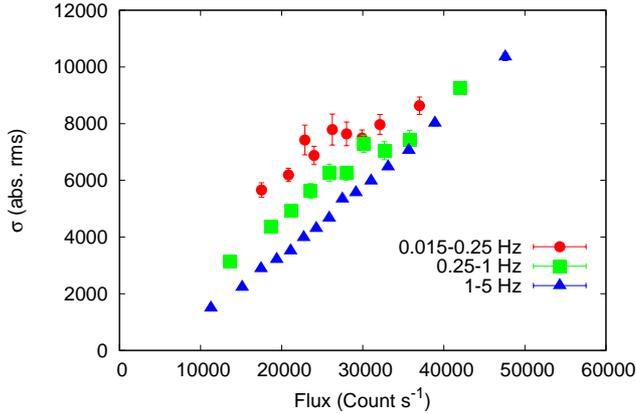}
 	\caption{Rms-flux relation evaluated in 3 different frequency ranges (0.015-0.25 Hz, 0.25-1 Hz, 1-5 Hz) for the IR light curve. The larger error bars for the low frequency data are due to a smaller number of segments than for the higher-frequency ranges.}
 	\label{fig:rms-flux}
 \end{figure}

 
 The relation for three different frequency ranges (0.015-0.25 Hz, 0.25-1 Hz, 1-5 Hz) is shown in Fig. \ref{fig:rms-flux}. A relatively tight linear relation is evident on all probed time scales. The larger uncertainties for the low-frequency measurements are due to the fewer number of segments with which the relation was computed. A linear fit was performed to quantify the parameters of the trend: the fitted function has the usual form $\sigma=k\cdot(F-C)$, where $\sigma$ is the absolute rms and $F$ is the IR flux. Values from the fits are reported in Tab. \ref{tab:sigma_flux_ir_fit}: at low frequencies the value of the intercept is < 0. A similar behaviour has been already observed in the optical \cite{gandhi2009}, and in the X-rays \citep{heil2012,gleissner2004}. From these works it emerged that a complex relation is present between intercept and the gradient, and that therefore the intercept cannot be interpreted simply as a constant flux component.
     \begin{table}
     	
     	\begin{center}

     		\ \begin{tabular}{ccc}
     		
     		\hline
     		Range [Hz] & $k$ [abs. rms/ct s$^{-1}$]& $C$ [$10^3$ ct s$^{-1}$]\\ \hline 0.015-0.25 & 0.15  $\pm$ 0.02 & -21 $\pm$ 3 \\ 0.25-1 & 0.229 $\pm$ 0.004 & -0.9 $\pm$ 0.5 \\ 1-5 & 0.238 $\pm$ 0.004&  5.7   $\pm$ 0.4\\ \hline
     		\end{tabular}	 
     		\caption{Parameters from the evaluation of a linear fit on the measured IR rms-flux relation. The fitted function is $\sigma= k(F-C)$. }
     		\label{tab:sigma_flux_ir_fit}
     	\end{center}
     \end{table} 
 
\begin{figure}
 	\includegraphics[width=\columnwidth]{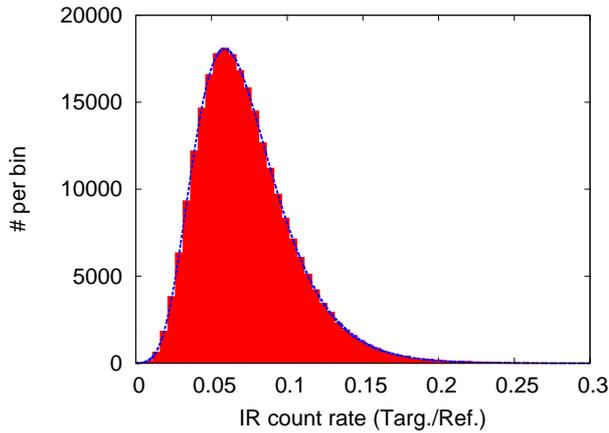}
 	\caption{Binned distribution of the IR count rate. The blue dashed line shows the best log-normal fit.}
 	\label{fig:log}
 \end{figure}

Given the presence of an rms-flux relation we also investigated the distribution of the fluxes. The histogram of the IR count rate is shown in Fig. \ref{fig:log}: a skewed distribution is clearly present. We also performed a fit to test whether the distribution is consistent with being log-normal. Even though the log-normal distribution seems to approximates the measured histogram the $\chi^2$ test was not successful ($\chi^2$/d.o.f.=96/52). The presence of excess of residuals is common in log-normal flux distributions fitting and it is usually interpreted with the presence of non-stationarity \citep{uttley2005,gandhi2009,edelson2013,scaringi2015}. 
%
%
%

 \section{Discussion}
 
\label{discussion} 
   
 \subsection{Fast flux-flux  correlation}
 \label{disc-flux}
In a given spectral state the source fluxes observed at different wavelengths are expected to be correlated, with a slope that depends on the different dependence of the emission processes at play on the accretion rate. Thus, the measurement of a flux-flux correlation can provide us with key information on the emission processes, in turn offering insights on the physical mechanisms with which matter accretes into (and is ejected from)  compact objects. For example, given a set of reasonable assumptions on the radio emitting medium (see \cite{coriat2011} for details), a radio/X-ray correlation with a power law index depending on the radiative efficiency of the X-ray emitting medium is expected. Indeed, the measured power law index of $\sim 0.6$ observed in the hard state of several BHTs is nicely consistent with the slope expected for a radiatively inefficient X-ray emitter, while a radiatively efficient flow would cause a steeper index of 1.4, as is indeed observed in some neutron stars and in a subset of BHTs (e.g. \cite{coriat2011,tudor2017}).

The X-ray flux has been found to correlate also with the emission at shorter wavelengths \citep{homan2005,russell2006,coriat2009}: in the detailed study made by \cite{coriat2009} for GX 339-4 during its low-hard state, the authors found - using daily X-ray (3-9 keV) and IR (H band, central wavelength: 1.62 $\mu$m, width: 0.27$\mu$m) observations - a broken power law correlation with a break at $L_{X}=5 \times 10^{-11} $ erg  s$^{-1} $cm$^{-2}$. Below the break the measured spectral index is $\beta^*=0.68 \pm 0.05$, while above the slope is $\beta^*=0.48 \pm 0.01$. \cite{coriat2009} interpreted the broken power law correlation with two possible scenarios: in the first one, the X-ray emission originates from a synchrotron self-Comptonization (SSC) process, while the IR emission is identified as optically-thin (frequencies above the break) or optically-thick (frequencies below the break) synchrotron emission from a jet. In the second scenario, the X-ray emission comes from a radiatively inefficient hot inflow (an ADAF for high luminosities which emitted mainly bremsstrahlung radiation at low luminosities) , while the IR emission arises from the optically thick regions of the jet (an optically thin origin for the IR emission is excluded in this case).

The mean 3-9 keV flux in our observation is $L_X=1.2 \times 10^{-10}$ erg  s$^{-1} $cm$^{-2}$, thus we are well above the break. However two of the three slopes we find are significantly different from the one measured at similar luminosities on longer timescales by \cite{coriat2009}. This may suggest that the process responsible for the luminosity variations on long timescales is different from the one driving the variations on short timescales. For instance, the long-term evolution of the jet luminosity could be scaling with accretion rate, while variability on shorter timescales could be driven by internal processes such as turbulence or shocks, which may modify the correlation between the jet emission and the mass accretion rate. We note however that \cite{coriat2009} did not include the 2008 outburst of GX 339-4 in their study. It is thus possible in principle that the 2008 outburst has a different IR/X-ray correlation slope than the 4 outbursts studied by \cite{coriat2009}. Nevertheless, while \cite{corbel2013} showed that the slope of the radio/X-ray correlation can change from outburst to outburst, which results in a large dispersion in the overall radio/X-ray correlation, the same authors note that a much smaller dispersion is observed in the IR/X-ray correlation, which points toward a rather stable relation from outburst to outburst.
%

Additionally, and independently of this, we also find possible hints for the IR/X-ray correlation slope, as measured on short timescales, to vary during our very dataset. 
It is known \citep[][and references therein]{coriat2009,coriat2011} that the slope depends strongly on whether the IR emission arises from the optically thick or optically thin regions of the jet. If the location of the break changes slightly, the variable spectral energy distribution (SED) in that frequency range would lead to a different slope in the flux-flux correlation. Namely, when the self-absorption break moves slightly toward longer wavelengths, the contribution to the observed IR flux from the optically-thin regions of the jet increases, steepening the IR/X-ray correlation. Vice versa, when the break moves toward higher frequencies, the IR/X-ray correlation flattens as a consequence of the increased IR contribution from optically-thick synchrotron.  Therefore, a possible explanation for the observed variable slope is that the IR emission has both optically thin and optically thick contributions, which would imply that the self-absorption break is at near-IR wavelengths.{ We note that a variable self-absorption jet break has been already reported for GX 339-4 \citep{gandhi2011} on timescales consistent with our result. The frequency range where the break was observed moving, is partially consistent with the one probed in this paper (from $\approx$ 20 $\mu$m, to $\approx$ 3 $\mu$m). Given that GX 339-4 does not display a significant evolution in hardness during the hard state\citep{buxton2012}, this is in agreement with the recent indications of a correlation between the jet break and the spectral hardness \citep{russell2013,koljonen2015}. }Significant variability of the IR spectrum has also been observed on timescales of $\approx$ 20 s \citep{Rahoui2012} which could indicate variations of the jet physical conditions. In addition, in principle the slopes could be affected by a thermal component. At shorter wavelengths, the accretion disc is known to contribute, producing a bluer spectrum in the optical compared to the IR \citep{Corbel2002,gandhi2011,CadolleBel2011,Rahoui2012}. However, at the $K_S$ band wavelength the disc contribution is minimal throughout the hard state, except possibly at the lowest fluxes \citep[see e.g. fig. 5 in][]{homan2005}, so the disc is unlikely to be responsible for the observed slope changes.

Additionally, a prediction of this scenario is that the IR linear polarisation should slightly decrease when the correlation is flatter, as the optically-thick synchrotron emission has an intrinsically lower degree of polarisation (by a factor of two) than the optically-thin synchrotron emission. We also searched for other observables that could vary during our observation, correlating with the flux-flux slope (i.e. PSD, coherence, lags). We found none, as the variability at both wavelengths appears to be stationary on the relevant timescales. Simultaneous multi-band high time resolution observations, to build a time-resolved SED, will help to test this scenario.

  
 
\subsection{Fourier Analysis}
 \label{disc-coh}


The observations reveal a highly variable source, both in X-rays and IR, with a 0.016-8 Hz fractional rms of 20.9 $\pm$ 0.1 per cent  and  13.20 $\pm$0.05 per cent respectively. While the X-ray PSD is consistent with a rather simple monotonic trend, the IR has a more complex PSD. In particular as we can see from Tab. \ref{tab:fit_lorenzt}, the main difference between the two bands is at frequencies higher than $\approx$ 2 Hz, where the IR PSD shows an excess (Fig. \ref{fig:psd}, Lower panel), followed by a break: this is due to the presence in the IR PSD of a strong Lorentzian component (normalization and width are higher than the ones measured for the low frequency component) peaking at $\approx$0.4 Hz with the addition of a third, weaker component centered at $\approx$ 1 Hz. A similar behavior was observed with simultaneous X-ray/optical observations \citep{gandhi2010} during the 2007 outburst of GX 339-4. The PSDs measured by \cite{gandhi2010} showed a shape and a parametrization similar to that of our dataset, with the higher-frequency component in the X-ray PSD seemingly replaced by two components in the optical PSD: a strong one, at lower frequencies than the X-ray \lq\lq missing" component, and a second weaker one at higher frequencies.  It is interesting to note that both the X-rays and the optical power spectra in the 2007 observations have characteristic frequencies a factor of $\sim 2-3$ higher than in the 2008 (X-ray and infrared) power spectra. This is evident when looking at the the optical PSD, which shows a break at higher frequencies (Optical $\nu_{\rm Break}\approx$ 3 Hz, IR $\nu_{\rm Break}\approx$ 1 Hz). We emphasize here that a similar excess followed by a break is also found in the PSD from O-IR synthetic signal generated by internal shocks models \citep{malzac2014}.  Damping of the high frequency variability is mainly due to physical size of the emitting region in the jet  \citep{malzac2014}. For this reason the model also predicts naturally  a break frequency which decreases as a function of the electromagnetic wavelength.   The measure of the energy dependence of the break in the PSD would be a key result in the study of jets, allowing constraints to be put on their physical size. However, as the dataset in \cite{gandhi2010} is not simultaneous with the one presented in this paper, it is not yet possible to draw further conclusions.




The X-ray and IR signals appear to be strongly correlated, with an intrinsic coherence of up to 90  on the longest timescales and decreasing below 30\% at frequencies above 1 Hz. The time lags are consistent with being constant as a function of frequency, pointing toward a scenario of simple propagation, in which input signal variability appears delayed in the output. However, the expected coherence from a simple time delay process (i.e. expressed mathematically as the convolution of the input signal with an approximately symmetrically peaked narrow impulse response function\footnote{In a linear system, the impulse response function (IRF) describes how the output signal is related to the input. Mathematically, if we have: $y(t)=\int r(t-\tau)x(\tau)d\tau$ (where $y(t)$ and $x(t)$ are respectively the input and output signals), the term $r(t-\tau)$ is known as the IRF. In the Fourier domain, the equation becomes:  $Y(\nu)= R(\nu)X(\nu)$, and $R(\nu)$ is called the \lq\lq transfer function\rq\rq. For further details, see also \citet{jenkins,nowak1999,uttley2014}.}) is constant in frequency. This is not consistent with the slow and smooth decreasing trend we observe in the coherence.  Therefore there must be some mechanism which affects the correlation between the two signals, without damping dramatically the IR variability.

Given the above considerations we suggest that a possible mechanism to explain such a trend could be a time-dependent impulse response function, which varies around an average value. We know that the coherence can be considered as a measure of the \lq\lq stability\rq\rq \space of the correlations across the light curve \citep{nowak1999}. Thus, a transfer function whose parameters are time dependent can reduce the coherence at high frequencies without influencing the lags too much. Moreover on long timescales the effect of a randomly variable lag would naturally average out, resulting, as a matter of facts, as a standard linear process. Of course, more complex scenarios are clearly possible, for example with the fine-tuned addition of uncorrelated components or further non-linear processes. A detailed description of the various solutions is beyond the aim of this paper and will be tested in future works. However, we note that a somewhat similar behaviour is in principle already at play in the internal-shocks model, where the IR is emitted in a very broad zone of the jet \citep{malzac2013}.
 

\subsection{rms-flux relation}

The presence of the rms-flux relation in most of the known accreting sources indicates that it is a fundamental property of the accretion process. From a mathematical point of view, a correlation between the mean of a quantity and its root mean square is expected whenever the distribution is not symmetric \cite{uttley2005}; therefore the linear rms-flux relation can be seen as a direct consequence of the log-normal distribution of the fluxes found in most accreting sources. As a log-normal distribution is usually associated with a multiplicative process, it is generally believed that the rms-flux relation emerges from the coupling of the variability on different timescales, which leads to non-linearity in the signals generally observed in these systems. From a physical point of view, the described properties are naturally explained by the \lq\lq propagating fluctuation\rq\rq \space model \citep{lyubarskii1997}.{ This model predicts that longer time-scale accretion rate variations are produced at larger radii in the accretion inflow .}  As variations propagate inwards, they combine multiplicatively with the locally produced variations, so that in effect, faster variations are modulated by slower ones. This scenario is also consistent with the hard X-ray lags found in different studies \citep{miyamoto1992,nowak1999,uttley2011}. {Our discovery of a linear rms-flux relation in the IR jet emission from an X-ray binary, means that this correlation is present also in a component where matter is outflowing from the source. 
Therefore, as the above propagation scenario cannot work in the outflow, either the rms-flux relation is transferred from the inflow, or it is originated by another mechanism in the outflow.
}

Evidence for a linear rms-flux relation has already been found in the light curves from other emitting components in other energy bands; however, the reason why the rms-flux relation emerges in these cases is still unclear. \cite{gandhi2009} first measured a linear rms-flux relation in the optical for XRBs. Based on the results from \cite{malzac2004} and \cite{zhang}, these authors suggested that the rms-flux relation in an optically emitting corona or jet could simply derive from the non-linear variability of the input signal (i.e. the X-rays). This includes the fastest frequency range extending above 1 Hz, where an optically-thin relativistic jet seems to be the dominant contributor to the optical emission \citep{gandhi2010}. It is easy to understand that the possibility of conserving the input nonlinearity can in principle work for a jet as well as for a corona. On a completely different black-hole mass scale, \cite{edelson2013} found the first evidence of an rms-flux correlation in the optically thin synchrotron emission from an extragalactic relativistic jet, using Kepler optical data of a blazar. Those authors recalled that an rms relation is a natural consequence of the \lq\lq mini-jet-in-a-jet" model \citep{biteau2012}, in which shocks with random orientation in subregions of the jet generate the observed variability.

The two reported examples represent very well the conceptual problem of the nature of the rms-flux relation in outflowing components: is the presence of an rms-flux relation just a conservation of variability properties from the inflow to the outflow? Or is it an intrinsic property of the emission process in the outflowing component itself?
Some clues may be contained in the frequency dependence of the rms-flux relation.  For example, as the frequency range used to measure the rms increases, the intercept on the flux axis systematically increases from negative to positive values.  This behaviour cannot be explained by a constant flux component which would affect all frequencies in the same way - instead, it implies that the flux offset is linked to a variable component which contributes differently to the rms on different time-scales and does not follow the simple linear rms-flux relation.  It is tempting to associated these distinct components with distinct components in the rms which are either coupled to the disk (dominating the linear relation) or intrinsic to the jet (dominating the intercept component).  However, this interpretation conflicts with the fact that the rms of the intercept component of rms is weakest at the higher frequencies ($\nu>1$Hz), where the coherence is low and therefore where we expect any component intrinsic to the jet but not to the accretion variability (which drives the X-ray variations) to be strong, not weak.  

There are a number of possible explanations for these contrasting results.  One is that the low coherence at high frequencies is not due to independent variability between X-rays and IR. Rather it tells us that there is either a variable impulse response or some non-linear transform relating the two bands (e.g. see discussion in \citealt{vaughan1997}), such that the coherence is intrinsically low.
Alternatively, the rms-flux relation may indeed be intrinsic to the variability generated within the jet. As mentioned in section \ref{disc-flux}, optically thick synchrotron radiation could be a significant part of the emission, therefore shocks could give a significant contribution to the observed variability (see also \cite{jamil2010,malzac2013,malzac2014}).  Therefore, this may imply that the shocks themselves can give rise to an rms-flux relation.  Indeed, synthetic lightcurves from internal shocks models already show typical non-linear features \citep{malzac2014}, and from preliminary analysis it is possible to see that a linear rms-flux relation is present (Vincentelli et al. in prep.); however, also in this case, the origin of such non-linearity is still an open question. Further analysis on the output from internal shocks models will help to shed new light on this problem, helping constraining the jet emission and re-acceleration mechanisms.
 
\section{Conclusions}
\label{conclusions}
We presented the first complete characterisation of the simultaneous X-ray and IR sub-second variability for the BHT GX 339-4. We summarise our main results as follows:

\begin{itemize}

\item We find a flux-flux power law correlation on timescales as short as 16 seconds, which is steeper than the one reported before on longer timescales (days). The slope appears to be variable on $\sim$hours timescales, consistent with the IR emission being a variable combination of optically-thin and optically-thick jet emission, perhaps as a consequence of a variable jet break located at IR wavelengths. 
\item The high and smoothly decreasing coherence as a function of frequency suggests that a scenario with only a simple time delay for the measured frequency-independent infrared lag of 100 milliseconds can be excluded. Additional fine-tuned uncorrelated IR variability would be needed to explain the observed phenomenology. We suggest instead that some \lq\lq dissipation" of the correlation must be involved, for example in terms of a \lq\lq flickering" impulse response function.

\item We measured for the first time the presence of a linear rms-flux relation in securely-identified jet IR emission: the low coherence measured on short timescales between the IR and the X-ray variability might suggest either that the accretion and jet emission are coupled but by a non-linear or time-variable transform, or that the IR rms-flux relation is not transferred from the inflow to the jet, but is generated within the jet emission processes.

\end{itemize}

The analysis presented here is a further confirmation of how powerful multi-wavelength high time resolution observations are to study the disc-jet interaction in XRBs. Furthermore, these results underline the need for theoretical models to include variability as a key ingredient, so as to make predictions for the large number of observables. Further observations, as well as a deeper analysis of the output lightcurves from existing and future jet variability models, are needed to improve our understanding of the physics of these systems.

\section*{Acknowledgements}

FV thanks Julien Malzac, Chris Done, Simon Vaughan and Michiel van der Klis for insightful discussions on the physical implications of the coherence and on the nature of the IR rms-flux relation.  Python package \lq\lq pyLCSIM" developed by Riccardo Camapana (http://pabell.github.io/pylcsim/html/code.html) was used to simulate lightcurves. FV acknowledges support from MPNS COST Action MP1304 and AHEAD Trans-National Access. PG acknowledges STFC for support (grant reference ST/J003697/2). BDM acknowledges support from the Polish National Science Center grant Polonez 2016/21/P/ST9/04025. LS acknowledges support from ASI-INAF agreement I/37/12/0

\end{document}